\documentclass[12pt]{article}
\usepackage{amsxtra,amssymb,amsthm,amsmath,latexsym}

\theoremstyle{plain}

\newcommand{\refS}[1]{Section~\ref{S:#1}}

\def\R{{\mathbb R}}

\def\C{{\mathbb C}}

\def\calF{{\mathcal F}}
\def\calM{{\mathcal M}}

\def\tildeC{\widetilde C}
\def\oH{{\overset{\circ}{H}}}
\def\oH1{{\overset{\circ}{H}\kern-.02in{}^1}}
\def\l{\ell}

\def\bee{\begin{equation*}}
\def\eee{\end{equation*}}
\def\be{\begin{equation}}
\def\ee{\end{equation}}

\begin{document}
\title{Distribution of particles which produces a desired radiation pattern}

\author{A.G. Ramm\\
 Mathematics Department, Kansas State University, \\
 Manhattan, KS 66506-2602, USA\\
ramm@math.ksu.edu,\\ fax 785-532-0546, tel. 785-532-0580}

\date{}
\maketitle\thispagestyle{empty}

\begin{abstract}
\footnote{PACS 03.40.Kf\, MSC 35J05, 35J10, 70F10, 81U40, 35R30   }
\footnote{key words: acoustic scattering, many-body problem, 
nanotechnology, inverse problems  }

A method is given for calculation of a distribution of small particles, embedded in a medium, so that the resulting medium would have a desired radiation pattern for the plane wave scattering by this medium.
\end{abstract}


\section{Introduction}\label{S:1} Let $D_0\subset \R^3$ be a
bounded domain, $k_0$ be the wave number in $D_0$ and $k<k_0$
be the wave number in $D'_0=\R^3\setminus D_0$. We assume
that $D_0$ is a homogeneous medium, so that $k_0$ is a constant.
This assumption can be weakened: we may assume that $k_0=k_0(x)$
is a known function.  Let
$D_m$ be a particle, $d_m$ be its diameter, $1\leq
m\leq M$, $M$ is the number of small particles, $a=\max_{1\leq m\leq 
M}\frac{d_m}{2}$ is an estimate for the  radius of a small particle. We 
assume that $k_0a\ll 1$, and then $ka\ll 1$, i.e., particles are small, 
that $d\gg a$, where
$d=\min_{j\not= m} dist (D_m,D_j)$ and that the particles are 
acoustically
soft, i.e., $u=0$ on $S_m$, the boundary of $D_m$. Denote
$U^M_{m=1} D_m:=U$, $R^3\setminus U=V$,
 $\partial U$ is the boundary of $U$, $C_m$ is the
electrical capacitance of a perfect conductor with the shape of $D_m$. 

The
Inverse Problem  is: 

{\it Can one distribute many small
particles in $D_0$ so that a plane wave
$u_0:=e^{ik\alpha\cdot x}$, (where $\alpha$ is a given unit
vector, $\alpha\in S^2$, $S^2$ is a unit sphere), scattered
by $D_0$, would produce a desired radiation pattern
(scattering amplitude) $A(\alpha',\alpha)$?}

The acoustic pressure $u$ solves the problem (1)-(3):
\be\label{E1}
  (\nabla^2+k^2-q(x))u=0 \hbox{\ in\ } V, 
  \qquad q(x)=\begin{cases} k^2-k^2_0 & \hbox{\ in\ } 
D_0\\
                                                0 & 
\hbox{\ in\ } D'_0,\end{cases}
\ee
\be\label{E2} u=0 \hbox{\ on\ } \partial U, \ee
\be\label{E3} 
  u=e^{ik\alpha\cdot x} + A(\alpha',\alpha) \frac{e^{ikr}}{r}+o\left(\frac{1}{r}\right), 
  \qquad r:=|x|\to\infty,
  \qquad \alpha'=\frac{x}{r},
\ee
and the coefficient $ A(\alpha',\alpha)$ is called the scattering 
amplitude or radiation pattern.

Let $G=G(x,y)$ solve the problem 
\be\label{E4}
  [\nabla^2+k^2-q(x)]G=-\delta(x-y)\hbox{\ in\ }\R^3,
  \qquad r\left(\frac{\partial G}{or}-ikG\right)=o(1) 
  \qquad r\to\infty.
\ee
This $G$ exists, is unique, and solves the integral equation
\be\label{E5}
G(x,y)=g(x,y) - (k^2-k^2_0) \int_{D_0} g(x,z) G(z,y)\,dz,
\qquad g=g(x,y):=\frac{e^{ik|x-y|} }{4\pi |x-y|}.
\ee
For brevity, we drop the $k$-dependence in $g$, $G$ and other 
functions below.
One may consider $G(x,y)$ known since $D_0$ and $k_0$ are known.

Let us look for the solution to \eqref{E1}--\eqref{E3} of the form
\be\label{E6}
  u=U_0(x,\alpha) + \sum^M_{m=1} \int_{S_m} G(x,t)
\sigma_m(t)\,dt,
\ee
where $\sigma_m(s)$ are unknown functions and $U_0(x,\alpha)$ is the 
scattering solution corresponding to the potential $q(x)$ in the absence 
of small bodies, i.e., in the whole space. For arbitrary $\sigma_m$ the 
right-hand side of \eqref{E6} solves 
equation \eqref{E1} (because $G$ solves \eqref{E4}), satisfies the 
radiation condition \eqref{E3}, and
\be\label{E7}
  A(\alpha',\alpha)= A_q(\alpha',\alpha)+\frac{1}{4\pi} \sum^M_{m=1} 
\int_{S_m} U_0(s,-\alpha') 
\sigma_m(s)\,ds,
\ee
where the scattering solution $U_0(s,\alpha)$ can be defined by the 
formula (\cite[p.232]{R1}):
\be\label{E8}
  G(x,s)=\frac{e^{ikr}}{4\pi r} U_0(s,\alpha) + 
o\left(\frac{1}{r}\right),
  \qquad r:=|x|\to\infty,
  \qquad \alpha=-\frac{x}{r}.
\ee
Formula \eqref{E8} was proved in  \cite[p.46]{R3},
where it was shown that $U_0(x,\alpha)$ solves the problem
\be\label{E9}
  [\nabla^2+k^2-q(x)] U_0(x,\alpha)=0\hbox{\ in\ } \R^3, 
  \qquad U_0 =e^{ik\alpha\cdot x}+A_q(\alpha',\alpha)\frac{e^{ikr}}{ 
r}+o\left(\frac{1}{r}\right), 
\ee
where $r:=|x|\to\infty$ along the direction $\alpha'$.
The function $U_0$ can also be considered known, and $G$ can be 
considered as Green's function for the Schr\"odinger equation \eqref{E9}.
The right-hand side of \eqref{E6} solves problem \eqref{E1}--\eqref{E3}
if and only if $\sigma_m$ are such that boundary condition \eqref{E2} is 
satisfied:
\be\label{E10}
  -U_0(s,\alpha)-\sum_{m\not= j} \int_{S_m} G(s,t) \sigma_m(t)\, dt
  = \int_{S_j} G(s,t)\sigma_j(t)\,dt,
  \qquad s\in S_j,
  \qquad 1\leq j\leq M.
\ee
So far the smallness of the particles was not used. If $ka\ll 1$, then equation \eqref{E10} can be
simplified:
\be\label{E11}
  -U_0(s_j,\alpha) -\sum_{m\not= j} G(s_j,t_m) Q_m
  =\int_{S_j} g_0(s_j,t)\sigma_j\,dt,
  \quad 1\leq j\leq M,
  \quad Q_m:=\int_{S_m} \sigma_m\,dt,
\ee
where $s_j\in S_j$ is any point in $S_j$, and  $g_0=\frac{1}{4\pi 
|x-y|}$.
In equation (11) we have used two approximations. The first one is
\be\label{E12}
  G(s,t)\approx g(s,t), \qquad s,t\in S_j.
\ee
This approximation is justified by \eqref{E5} when $k|x-y|\ll 1$, 
because the term $g(x,y)$ is the main term on the right-hand side of 
\eqref{E5} as $x\to y$, and $|s_j-t|\ll 1$.
The second one is $g\approx g_0$ if $k|s-t|\ll 1$, and
the error of this approximation is $O(ka)$, so this approximation is
also justified because $ka\ll 1$.

Equation \eqref{E11} is equation for the charge 
distribution $\sigma_j$ on the surface
$S_j$ of a perfect conductor charged to the potential 
$-U_0(s_j,\alpha)-\sum_{m\not= j} G(s_j,t_m)Q_m$.
Therefore the total charge $Q_j:=\int_{S_j} \sigma_j dt$ 
on the surface $S_j$ can be calculated by the formula
\be\label{E13}
  Q_j=C_j \left(-U_0(s_j,\alpha) - \sum_{m\not=j} G(s_j,t_m)Q_m\right),
  \qquad 1\leq j\leq M, \ee where $C_j$ is the electrical capacitance of
the perfect conductor $D_j$.  In \cite[p.385]{R1} formulas for calculation
of $C_j$ with arbitrary desired accuracy are given. Equation \eqref{E13}
is a linear algebraic system for finding unknown $Q_j$, $1\leq j\leq M$.

Consider the limiting case $M\to\infty$ of the distribution of particles
in $D_0$. Define $C(y)$ as follows: 
\be\label{E14}
  \int_D C(y)\,dy=\lim_{M\to\infty} \sum_{D_m\subset D} C_m,
\ee
where $D\subset D_0$ is an arbitrary subdomain of $D_0$.
The above definition means that $C(y)$ is the limiting density of 
the capacitances of the small particles in $D$.
Formula \eqref{E6} shows that the self-consistent field $u_e$ can be defined as
\be\label{E15}
  u_e(x,\alpha)=
  \begin{cases} U_0(x,\alpha) +\sum^M_{m=1} G(x,t_m)Q_m, & 
min_{1\leq m \leq M}|x-t_m|\gg a,\\
                         U_0(x,\alpha) +\sum_{m\not= j} G(x,t_m)Q_m, & 
|x-t_j|\sim a.
  \end{cases}
\ee  
Therefore, defining $u_e$ we neglect the influence of any fixed
single small particle on the field.
This is justified when $M\to\infty$.
Using  \eqref{E13} and \eqref{E15} one gets
\bee
  u_e(x,\alpha)= U_0(x,\alpha)- \sum_m G(x,t_m)C_m u_e(t_m,\alpha),
\eee
and in the limit $M\to\infty$ one obtains the equation:
\be\label{E16}
  u_e(x,\alpha)= U_0(x,\alpha) -\int_{D_0} G(x,y) C(y) u_e(y,\alpha)\,dy,
\ee
where $C(y)$ is defined in \eqref{E14}. Equation \eqref{E16} is equivalent to the 
Schr\"odinger scattering problem:
\be\label{E17}
  [\nabla^2+k^2-q(x)-C(x)] u_e=0 \hbox{\ in\ } \R^3,
\ee
where $q(x)$ is known, the function $u_e$ has the following asymptotics
\be\label{E18}
  u_e=e^{ik\alpha\cdot x} +A(\alpha',\alpha) \frac{e^{ikr}}{r}+o\left(\frac{1}{r}\right),
  \qquad r:=|x|\to\infty,
  \qquad \alpha'=\frac{x}{r},
\ee
and  $A(\alpha',\alpha)$ is the scattering amplitude at a fixed $k>0$.

Therefore the Inverse Problem, stated above, is reduced to inverse 
scattering problem of finding the potential 
$q(x)+C(x)$ from the knowledge of the corresponding fixed-energy 
scattering 
amplitude $A(\alpha',\alpha)$.

This problem was solved by the author 
(see \cite[Chapter 5]{R1}  and references therein).
In \refS{2}  we outline the author's 
algorithm for solving this inverse scattering problem.

If the potential $q(x)+C(x)$ is found and $q(x)$ is known, then $C(x)$ is 
found and one knows the density of the particle
distribution in $D$ which produces the desired radiation pattern
$A(\alpha',\alpha)$. Assuming that the particles are identical, one has 
$C_m=\cal C$, where
$\cal C$ is the electrical capacitance of one particle, and 
$C(x)=N(x)\cal C$,
where $N(x)$ is the density of particles, that is, the number of
particles per unit volume around point $x$. See \cite{R2} for the theory
of wave scattering by small bodies.

\section{Solution to inverse scattering problem}\label{S:2}

We follow \cite[p.264]{R1} and take $k=1$ without loss of generality.
Given $A(\alpha',\alpha)$ one finds 
$A_\l(\alpha):=\int_{S^2} A(\alpha',\alpha)\overline{ 
Y_\l(\alpha')}\,d\alpha'$,
where $Y_\l:=Y_{\l m}$ are the normalized spherical harmonics, so
that
\be\label{E19}
  A(\alpha',\alpha)=\sum^\infty_{\l=0} A_\l(\alpha) Y_\l(\alpha'),
  \qquad \sum^\infty_{\l=0} :=\sum^\infty_{\l=0} \sum^\l_{m=-\l}.
\ee
If the data  $A(\alpha',\alpha)$ are exact, then (\cite[p.262]{R1})
\be\label{E20}
  \max_{\alpha\in S^2} |A_\l(\alpha)|\leq O
   \left(\sqrt{\frac{b_0}{\l}} 
\left(\frac{b_0e}{2\l}\right)^{\l+1}\right),
   \quad |Y_\l(\theta')| \leq \frac{1}{\sqrt{4\pi}}\ 
\frac{e^{\kappa r}}{|j_l(r)|}
   \quad \forall r>0,\quad \theta\in \calM,
\ee
where $b_0>0$ is the radius of the smallest ball containing the domain 
$D_0$,
$\calM:=\{\theta:\theta\in\C^3, \theta\cdot\theta=1\}$, 
$\theta\cdot \omega:=\sum^3_{j=1} \theta_j\omega_j$, $\kappa=|Im\theta|$, 
$j_\l(r)$ is the spherical Bessel function. 
Fix an arbitrary $\xi\in\R^3$.
One can find (nonuniquely and explicitly) $\theta',\theta\in \calM$,
such that
$\theta'-\theta=\xi$, $\theta\to\infty$.
For example, if $\xi=te_3$, $t=|\xi|>0$,
(which can be assumed without loss of generality), then
$\theta'=\frac{t}{2}e_3 + z_1e_1+z_2e_2$, $\theta=-\frac{t}{2} 
e_3+z_1e_1+z_2e_2$,
and the condition 
\be\label{Estar}
  \frac{t^2}{4} + z^2_1+z^2_2=1, \quad z_1,z_2\in\C,\tag{$\ast$}
  \ee
implies $\theta,\theta'\in\calM$.
One may find many $z_1z_2\in\C$, such that \eqref{Estar} holds and $|z_1|\to\infty$.
For example, take $z_1=re^{i\varphi}$, $z_2=re^{-i\varphi}$, then
$r^2\sin(2\varphi)-r^2\sin(2\varphi)=0$, 
$r^2[\cos(2\varphi+\cos(2\varphi)]=1-\frac{t^2}{4}$,
so $r^2\cos(2\varphi)=\frac{1}{2}-\frac{t^2}{8}$.
One can take $r\geq |\frac{1}{2}-\frac{t^2}{8}|^{1/2}$ and find 
$\varphi$ such that $\cos(2\varphi)=\left(\frac{1}{2}-
\frac{t^2}{8}\right)\frac{1}{r^2}$.

In what follows we always assume
\be\label{E21}
  \theta'-\theta=\xi, \qquad \theta',\theta\in\calM, \qquad |\theta|\to\infty.
\ee
Because of \eqref{E20}, the series
\be  \label{E22}
  A(\theta',\alpha)=\sum^\infty_{\l=0} A_\l(\alpha) Y_\l(\theta'),
  \qquad \theta\in\calM,
\ee
converges absolutely and uniformly on compact subsets of $S^2\times\calM$.  

Fix positive numbers $b_0<b_1<b_2$ such that 
$D_0\subset B_{b_0}:=\{x:|x|\leq b_0\}$.
Note that the scattering solution $u(x,\alpha)$ for the scattering 
potential $q(x)+C(x)$ can be written explicitly in the region $|x|>b_0$:
\be\label{E23}
  u(x,\alpha)=e^{i\alpha\cdot x}
    +\sum^\infty_{\l=0} A_\l(\alpha) Y_\l(\alpha') h_\l(r),
  \qquad r:=|x|>b_0,
\ee
where $k=1$ and $h_\l(r)$ is the spherical Hankel function which is 
normalized
by the asymptotics
\be\label{E24}  h_\l(r)\sim \frac{e^{ir}}{r} \qquad r\to\infty.\ee
Let $\nu(\alpha)\in L^2(S^2)$. Consider the problem:  
\be\label{E25}   \calF(\nu)=\min,\ee
where
\be\label{E26}
  \calF(\nu):=\int_{b_1\leq|x|\leq b_2} 
  \left| e^{-i\theta\cdot x} \int_{S^2} u(x,\alpha)\nu(\alpha)\,d\alpha-1\right|^2dx.
\ee
The function $u(x,\alpha)$ in \eqref{E26} is defined in \eqref{E23},
and the minimization in \eqref{E25} is with respect to all $\nu\in L^2(S^2)$. 
One can prove (\cite[p.265 ]{R1}) that 
\be\label{E27}
\inf \calF(\nu):=d(\theta)\leq \frac{const}{|\theta|},
  \qquad \theta\in\calM, 
  \qquad |\theta|\gg 1, 
\ee
Let  $\nu(\alpha,\theta)$ be an arbitrary approximate solution  to \eqref{E25} in the following sense:
\be\label{E28}  \calF(\nu(\alpha,\theta))\leq 2 d(\theta).  \ee
For this $\nu$ define
\be\label{E29}
  \hat{C} :=-4\pi\int_{S^2} A(\theta',\alpha)\nu(\alpha,\theta)\,d\alpha,
\ee
where $A(\theta',\alpha)$ is defined in \eqref{E22}.

Let $\tildeC (\xi)=\int_{D_0} C(x) e^{-i\xi\cdot x} dx$,
where $C(x)\in L^2(D_0)$ vanishes in $D'_0$. 
Let \eqref{E21} hold.

The following theorem is proved by the author in \cite[p.266]{R1}.

\noindent{\bf Theorem.}\,\,
{\it Under the above assumptions one has}
\be\label{E30}  
|\hat{C}-\tildeC(\xi)|=O\left(\frac{1}{|\theta|}\right). \ee

\noindent{\bf Conclusion:} An algorithm is given for embedding many small
particles in a domain $D_0$ in such a way that the plane wave, scattered
by such domain, would have a desired radiation pattern
$A(\alpha',\alpha)$.

The algorithm consists of solving the inverse scattering problem, namely,
finding the potential $q(x)+C(x)$, vanishing outside $D_0$, from the
fixed-energy $(k^2=const>0)$ scattering amplitude $A(\alpha',\alpha)$. If
this potential is found, then $C(x)$ is found, and the small 
acoustically soft identical particles
should be distributed in $D_0$ with the density $N(x)={\cal C}^{-1} 
C(x)$, 
where
$\cal C>0$ is the electrical capacitance of a perfect conductor which has 
the
shape of the particle.


\begin{thebibliography}{1000} 

\bibitem{R1}
 Ramm, A.~G.~,
 Inverse problems, Springer, New York, 2005.

\bibitem{R2} 
Ramm, A.~G.~,
Wave scattering by small bodies of arbitrary shapes, World Sci. Publishers, Singapore, 2005.

\bibitem{R3}
Ramm, A.~G.~,
Scattering by obstacles, D.Reidel, Dordrect, 1986.



\end{thebibliography}
\end{document}